\documentclass[runningheads]{llncs}

\usepackage[T1]{fontenc}
\def\doi#1{\href{https://doi.org/\detokenize{#1}}{\url{https://doi.org/\detokenize{#1}}}}
\usepackage{listings}
\usepackage{amsmath}
\usepackage[utf8x]{inputenc}
\newenvironment{sloppypar*}
{\sloppy\ignorespaces}
{\par}
\usepackage{graphicx}
\usepackage{xcolor}
\usepackage{tikz}
 
\usepackage{amsmath}
\newcommand{\remove}[1]{}

\usepackage[linesnumbered,ruled,vlined]{algorithm2e}

\begin{document}
\title{Partially Disjoint $k$ Shortest Paths 
}
\titlerunning{Partially Disjoint $k$ Shortest Paths}
%
\author{Yefim Dinitz\inst{1}\and
Shlomi Dolev\inst{1} \and
Manish Kumar\inst{1}\and
Baruch Schieber\inst{2}}
\authorrunning{Dinitz et al.}
%
\institute{Ben-Gurion University of the Negev\and
New Jersey Institute of Technology\\ \email{\{dinitz,dolev@cs,manishk@post\}.bgu.ac.il,\\ sbar@njit.edu}}
\maketitle              

\begin{abstract}
A solution of the $k$ shortest paths problem may output paths that are identical up to a single edge. On the other hand, a solution of the $k$ independent shortest paths problem consists of paths that share neither an edge nor an intermediate node. We investigate the case in which the number of edges that are not shared among any two paths in the output $k$-set is a parameter.

We study two main directions: exploring \emph{near-shortest} paths and exploring \emph{exactly shortest paths}. 
We assume that the weighted graph $G=(V,E,w)$ has no parallel edges and that the edge lengths (weights) are positive. Our results are also generalized to the cases of $k$ shortest paths where there are several weights per edge, and the results should take into account the multi-criteria prioritized weight.
\end{abstract}

\section{Introduction}
Optimizing the cost of paths is a fundamental task in Computer Science and Operations Research. In many scenarios, there is a need to compute the second best or in general, the $k$ best alternatives to the optimal solution . The variety in the obtained $k$-set of best solutions can facilitate a choice of solutions due to other considerations (e.g., preferences of geographic locations or communication channels) among the close to optimal (or allowed budget) solutions. In some cases, the usage of all (or several) solutions from the set is preferable in order to allow the diversity of routing patterns while still being close to the optimal solution.

As an example of the applicability of the investigation of the set of $k$ near-optimal solutions consider genome exploration evaluation (e.g., for viruses variants). The distance relation, in this application, is defined as the probability of possible changes in the genome. The $k$ near-optimal solutions enable the tracing of the most probable paths from one variant to another or from one variant to many others.

One variant of this problem is finding $k$ independent shortest paths \cite{Eilam98}. This is a challenging computational task, where polynomial algorithms exist only for very limited cases. Thus, it leads to interesting theoretical questions that consider the relaxation of the independence constraint. Note that $k$ 1-edge independent shortest paths (i.e., paths that differ by at least one edge) can be computed in polynomial time. The increase from 1-edge independence to the entire path independence changes the complexity of the problem dramatically. We found polynomial algorithms for $k$ shortest paths that are $\ell$-edge  independent that can serve as an intermediate solution instead of totally independent paths.

Our algorithms on partially independent paths can be useful in a scenario where multiple objects or robots want to move (almost) collision-free~\cite{DBLP:journals/siamcomp/HopcroftW86,DBLP:journals/algorithmica/ErdmannL87}. Similarly, our algorithms may be applicable to multi-agent systems such as object transportation~\cite{DBLP:conf/iros/MataricNS95,DBLP:conf/iros/RusDJ95}, search and rescue~\cite{Jenn1997}, robot path reconfiguration\cite{DBLP:conf/focs/PapadimitriouRST94,DBLP:journals/corr/Jain21}, tasks spanning assemble~\cite{DBLP:journals/algorithmica/HalperinLW00}, evacuation~\cite{DBLP:conf/icra/RodriguezA10}, formation control~\cite{DBLP:conf/icra/SmithEH08}.

\noindent
\subsection{Related work}
One of the related directions in the literature is studying optimal sets of paths, where a bounded number of \emph{shared} vertices or edges is allowed. There, a set of paths is called optimal (shortest) if the \emph{sum of the path lengths} is minimum possible, where the length of a path is defined as the sum of the weights of all its contained edges. Observe that this objective does not imply any guarantee on the quality of each path. Guo et al.~\cite{DBLP:conf/ijcai/GuoDLHSH18} studied the problem of finding the shortest set of $k$ paths that are edge-disjoint and partially vertex disjoint. They considered the \emph{$\delta$-vertex $k$ edge-disjoint shortest path} ($\delta V-kEDSP$) problem, where at most $\delta$ vertices (besides $s$ and $t$) are shared by at least any two paths. 
As mentioned there, the $0V-kEDSP$ problem 
can be simply solved via min-cost max-flow.
For $k=2$ and any positive $\delta$, they solve the $\delta V-kEDSP$ problem in time $O(\delta m + n \log n)$, for a graph $G(V,E,w)$ with $n=|V|$ and $m=|E|$.
For general $k$, the problem is still open.
Similarly, the $k$ partially edge-disjoint path problem ($kPESP$) computes the shortest set of $k$ paths connecting $s$ and $t$ such that at most $\delta$ edges are shared by at least two paths. 
For $k=2$, Yunyun et al.~\cite{DBLP:conf/cocoon/DengGH18} introduced an exact algorithm with a runtime $O(mn \log_{(1+m/n)} n + \delta n^2 )$.
In the above works, $\delta > 0$ was referred to as the {\em disjointness} factor.

Chondrogiannis et al.~\cite{DBLP:conf/gis/ChondrogiannisB15} introduced the $k$ Shortest Paths with Limited Overlap \emph{(kSPwLO)} problem seeking to find $k$ alternative paths which are $(a)$ as short as possible and $(b)$ sufficiently dissimilar based on a user-controlled similarity threshold. Given a set of simple paths $P$ from a source $s$ to destination $t$ in an edge weighted graph $G(V,E,w)$, Chondrogiannis et al. called a path $p(s \xrightarrow{} t)$ an {\em alternative path} to $P$ if $p$ is sufficiently dissimilar to every path $p' \in P$. The similarity of two simple paths $p$ and $p'$ is determined by their overlap ratio:

\[
    Sim (p, p') = \frac{\sum_{(x, y) \in p \cap p'} w_{xy}}{w(p')},
\]
where  $w(p)$ is the length of a path $p$, and $p_i \cap p'$ denotes the set of edges shared by $p$ and $p$. The range of overlap ratio is $0 \leq Sim (p, p') \leq 1$, where $Sim (p, p') = 0$ holds if $p$ shares no edge with $p'$ and $Sim (p, p') =1$ if $p = p'$. Chondrogiannis et al. introduced two algorithms. The first is a baseline algorithm based on Yen's algorithm~\cite{10.2307/2629312}. The second algorithm, \emph{OnePass algorithm}, considers the overlap constraint in each expansion step while traversing the network.

In another work by Chondrogiannis et al.~\cite{DBLP:conf/edbt/ChondrogiannisB17} they considered the same similarity constraint and introduced the \emph{MultiPass (exact) algorithm} that traverses the network $k-1$  times and employs pruning criteria to reduce the number of potential alternative paths. The pruning criteria are the same as in the \emph{OnePass algorithm}.

Let $P$ be the given set of paths from source $s$ to destination $t$, and $p_i, p_j$ be two paths from source $s$ to some intermediate node $t'$. If $w(p_i) < w(p_j)$ and $\forall p \in P\ Sim (p_i, p) \leq Sim (p_j, p)$ holds, then the path $p_j$ cannot be the prefix of any of the shortest alternative paths to $P$. It takes $O(m + K \cdot n \cdot \log n)$ time, where $K$ ($K >> k$) is the number of shortest paths that have to be computed in order to cover the $k$ results of the $k$SPwLO query. In comparison to the OnePass algorithm, the MultiPass algorithm may have to construct all paths from $s$ to $t$, which results in higher time complexity, but experimental evaluation showed that MultiPass is much faster than OnePass.

Below, we list a sample of previous results on the disjoint paths problem. For a set of $k$ pairs of terminals in graph, the existence of $k$ vertex-disjoint paths connecting each pair of terminals, Robertson and Seymour~\cite{DBLP:journals/jct/RobertsonS95b} developed a $O(n^3)$ time algorithm for any fixed $k$ in their graph minor project. The problem of two disjoint shortest paths was first considered by Eilam-Tzoreff \cite{Eilam98}. Eilam-Tzoreff provided a polynomial-time algorithm for $k$= 2, based on a dynamic programming approach for the weighted undirected vertex-disjoint case. This algorithm has a running time of $O(|V|^8)$. Later, Akhmedov \cite{DBLP:conf/csr/Akhmedov20} improved the algorithm of Eilam-Tzoreff, whose running time is $O(|V|^6)$ for the unit-length case of the 2-Disjoint Shortest Path and $O(|V|^7)$ for the weighted case of the 2-disjoint shortest path. In both cases, Akhmedov \cite{DBLP:conf/csr/Akhmedov20} considered the undirected vertex disjoint shortest path. In recent past, Bentert et al.~\cite{DBLP:conf/icalp/BentertNRZ21} improved the result of Akhmedov \cite{DBLP:conf/csr/Akhmedov20}. In other work of Bérczi et al.\cite{DBLP:conf/esa/Berczi017} they showed that the undirected k-DSPP (disjoint shortest paths problem) and the vertex-disjoint version of the directed k-DSPP can be solved in polynomial time if the input graph is planar and $k$ is a fixed constant. Lochet~\cite{DBLP:conf/soda/Lochet21} shows that for any fixed $k$, the disjoint shortest paths problem admits a slicewise polynomial time algorithm.

\subsection{Preliminaries}

We propose to study variation of the  $k$ \emph{partially independent shortest paths}. The input to our problem consists of either directed or undirected weighted graph $G(V,E,w)$ with a length $w(e)$ associated with each edge $e \in E$, and an integer $k$. 

We assume that the graph has no parallel edges and no self loops. A (simple) path $p$ connecting two terminals $s,t \in V$ is a sequence of vertices: $s = v_0, v_1, \ldots , v_r = t$ such that all $v_i$’s are distinct and $(v_i, v_{i+1}) \in E$, for $i \in [0,1, \ldots, r - 1]$. Let $E(p)$ denote the set of edges in the path $p$, and $w(p)$ denote the length (weight) of path $p$, that is, $w(p) = \sum _{e \in E(p)} w(e)$. We also fix some independence measure as discussed below. Given this measure the goal is to find $k$ partially independent shortest paths connecting $s$ to $t$. (This can be extended naturally to the multiple terminal pairs case.)

Independence measure of shortest paths can be defined in several ways. In this note we view a path as a set of edges (or nodes) and consider the size of either the set difference or the symmetric difference of two paths as their independence measure.

As a warm-up, consider the 2 partially independent shortest paths problem and let the size of the set difference between the edge sets of the two paths be the independence measure of those paths.
By the pigeonhole principle, any collection of $m+1$ paths must include a pair of paths whose set difference is at least 2 (even assuming parallel edges).

Such $m+1$ near-shortest paths can be found using Yen's algorithm~\cite{10.2307/2629312}. This yields a solution of the 2 partially independent shortest path problem. 
for this variant of independence measure.
(Below, we give additional solutions for this independence measure.)

In this short paper we consider two variants of the partially independent paths problem. Section \ref{s:piksp} considers the case in which any subset of the $k$ shortest paths can be output in the solution 

and Section \ref{s:pisp} considers the case in which only strictly shortest paths qualify (and thus all qualified paths should be equal in length).

Note that all our results can be generalized to the multi-criteria prioritized weights case. That is, to the case where there are several weights per edge and any arbitrarily small amount of weight $i$ is more important than an arbitrarily big amount of weight $j$, for any $i < j$. This is possible by the reduction of this case to the case of a single weight provided in \cite{DBLP:conf/cscml/Dinitz21}.

\section{Partial Independence among $k$ Shortest Paths}
\label{s:piksp}
We consider the following approach: generate the near-shortest paths in the order from best to worst (e.g., using Yen's algorithm~\cite{10.2307/2629312}), and find a pair of paths with the highest independence measure among them. The independence measure that we consider is the size of the symmetric difference between either the edges or the nodes of the two paths.

\noindent
Our positive results are as follows (we provide no proof):
\begin{itemize}
    \item The first and the second near-shortest paths have at least three different edges and at least one different node.
    \item Among the first three near-shortest paths, there are two paths with at least four different edges and at least two different nodes.
    \item (A partly investigated conjecture) Among the first $O(n)$ near-shortest paths, there are two paths with at least six different edges and at least four different nodes.
\end{itemize}
For the last item, we do not have a full proof. We studied many cases,
and still are not sure that all possible cases are revealed. 

We also constructed two examples of graphs, where there is no pair of paths with a prescribed number of different edges/nodes among exponentially many first near-shortest paths.
The examples show that the effectiveness of the considered approach is bounded (even for planar graphs), since it implies that the worst-case time complexity is exponential in the prescribed distance between the two paths. 

Version (a) of Example 1 is a planar graph parametrized by $\bar q$, where for any $q \le \bar q$, among the first $\Omega((n/\bar q)^q)$ near-shortest paths, the maximal edge/node distances between two paths are $4q$ edges and $2q$ nodes. Example 1a has the following structure:
\begin{itemize}
    \item It has $n=\bar q (r+1)+1$ nodes, for  arbitrary $\bar q ,r \geq 1$. 
    \item Each path from $s$ to $t$ has $2 \bar q$ edges.
    \item The shortest path $p^*$ is composed of $\bar q$ pairs of consecutive edges, so that each one of those pairs can be replaced by any one out of $r-1$ other edge pairs.
    \item The edge weights are integers, and the maximal edge weight is about $r^ {\bar q}$.
\end{itemize}

\begin{figure}[!ht]
    \centering
    \includegraphics[page=1, width=\linewidth]{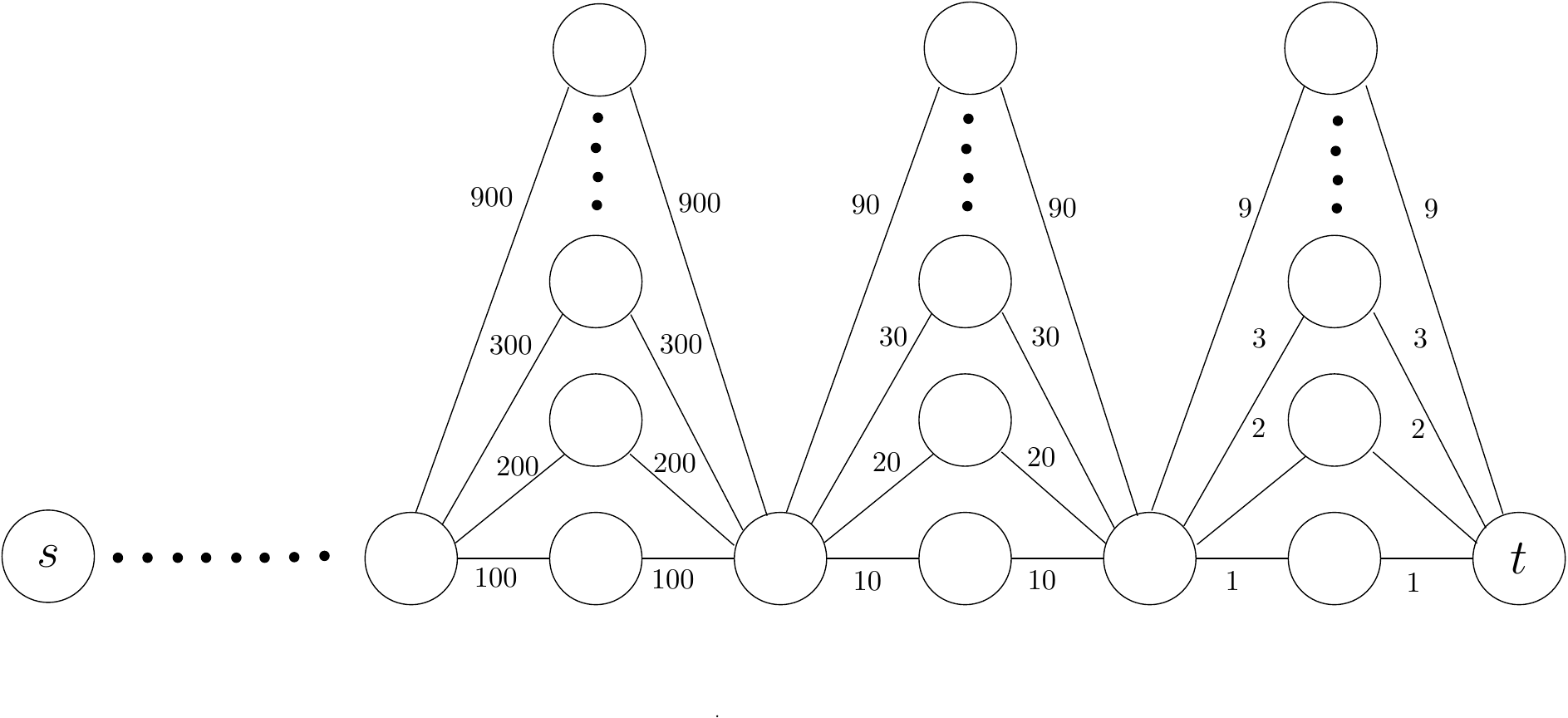}
    \caption{Example 1a with $r=9$ and $\bar q$ ``towers" between $s$ and $t$.}
    \label{fig:example}
\end{figure}

See Figure~\ref{fig:example} for an illustration of such a graph with the maximal edge distance of $4q$ and the node distance of $2q$ among all pairs of the first $9^q$ near-shortest paths, for any $q \le \bar q$.

\begin{figure}[!ht]
    \centering
     \includegraphics[page=1, width=0.95\linewidth]{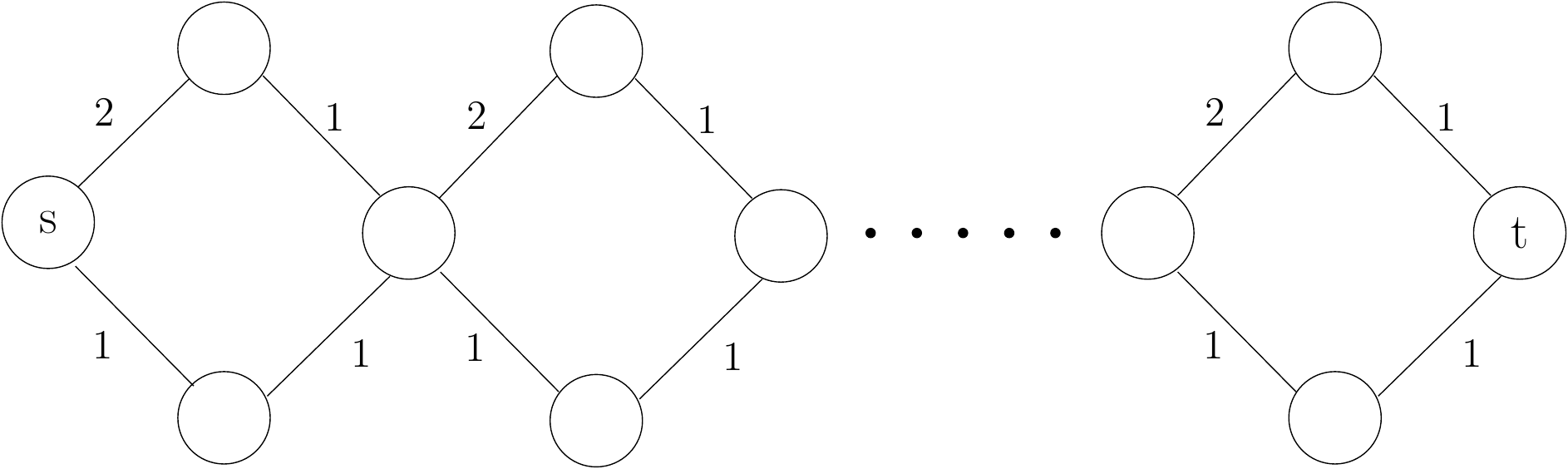}
    \caption{Example 1b.}
    \label{fig:example_1b}
\end{figure}

Version (b) of Example 1 has a simpler structure and eliminates the exponential edge weights by the cost of a bit weaker exponential properties. The graph on $n = 3n'+1$ nodes is also planar, constructed as a concatenation of $n'$ diamonds as in Figure~\ref{fig:example_1b}.
There, among the first $\sum_{i=0}^q C_{n'}^q = \Omega(((n'-q+1)/q)^q)$ near-shortest paths, the maximal edge/node distances between two paths are $8q$ edges and $2q$ nodes, for any $q \le \lfloor n'/2 \rfloor$.

Example 2, also parametrized by $\bar q$, shows how an arbitrary graph in a large graph class can be ``spoiled'' locally so that among the first $\sum_{i=1}^q \prod_{j=0}^{i-1} (\bar q - j) + 1 = \Omega((\bar q-q+1)^q)$ near-shortest paths, the maximal edge/node distances between two paths are $2q+2$ edges and $\min\{2q; \bar q\}$ nodes, for any $q \le \bar q$. 

\begin{figure}[!ht]
    \centering
    \includegraphics[width=\linewidth]{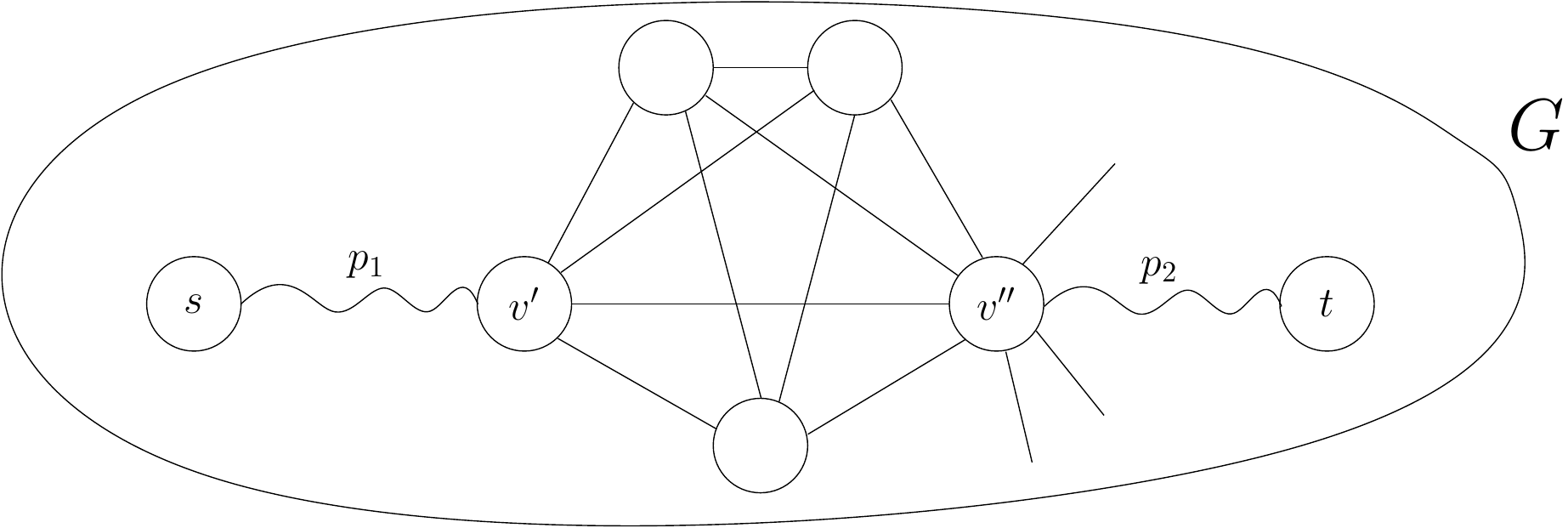}
    \caption{Example 2 with $\bar q = 3$.}
    \label{fig:example_2}
\end{figure}

Let $G_0=(V,E,w)$ be an arbitrary weighted graph with a \emph{single shortest path} from $s$ to $t$. Denote that path by $p^*$ and the second shortest path by $p^*_2$. We set $\epsilon = (w(p^*_2) - w(p^*))/(\bar q + 2) > 0$. 
Let $v$ be an arbitrary node at $p^*$, 
breaking $p^*$ into $p_1$ and $p_2$. We split $v$ into two nodes $v'$ and $v''$, so that $p_1$ finishes at $v'$, $p_2$ starts from $v''$, and all edges originally incident to $v$, except for that lying on $p_1$, are now incident to $v''$. We now add a complete graph, whose nodes are $v', v''$ and $\bar q$ new nodes and whose edges are of weight $\epsilon$ each, to the obtained graph. See Figure~\ref{fig:example_2} for an illustration.

Denote the resulting graph by $G$.
The shortest path from $s$ to $t$ in $G$ is $p_1 \circ (v',v'') \circ p_2$ of weight $w(p^*)+\epsilon$. The near-shortest paths in $G$, in order, are of the form $p_1 \circ p \circ p_2$, where $p$ goes over all paths from $v'$ to $v''$ of lengths 2, after that 3, and so on up to $\bar q + 1$ (note that the weights of all of those paths are strictly less than $w(p^*_2)$).
The claimed exponential properties of $G$ are easy to validate.

Note that the boundary case of Example 2 with $s=t$ is just a complete graph on $\bar q + 1$ vertices with all edge weights 1.

Another variant of Example 2 arises if we similarly ``spoil'' $G_0$ by inserting either version of Example 1 between nodes $v'$ and $v''$, with the edge weights proportionally decreased, instead of the complete graph as above. 
Importantly, if the original graph $G_0$ is planar, then the resulting graph $G$ will also be planar. The properties of this variant are also exponential, though a bit weaker than when using the complete graph.

\section{Partially Edge-Disjoint Exact Shortest Paths}
\label{s:pisp}

In this section, we study finding partially edge-disjoint paths among the \emph{$($exact$)$ shortest} paths. The restricted scope allows achieving new interesting results. 
Our main tool is the \emph{subgraph of shortest paths} $\tilde G$ introduced in \cite[Section 4]{DBLP:conf/cscml/Dinitz21}.\footnote{
This subgraph may be considered as a generalization of the layered network introduced in \cite{Dinic70}. A layered network $L=L(G)$ is a subgraph of the given unweighted graph $G$, such that the set of \emph{all shortest paths} from $s$ to $t$ in $G$ coincides with the set of \emph{all paths} from $s$ to $t$ in $L$.
}
For any graph $G=(V,E,w)$ and its nodes $s$ and $t$, its subgraph $\tilde G=\tilde G(s,t)$ is composed of the nodes and edges of all shortest paths from $s$ to $t$, keeping their weights. If $G$ is undirected, then its edges are directed along the shortest path(s) going through them, so the subgraph of shortest paths $\tilde G$ is always directed. 
Its main properties are as follows. We denote by $d(u,v)$ the distance (=the length of the shortest path) from node $u$ to node $v$.

\begin{figure*}[t]
    \centering
    \includegraphics[page=1, width=0.70\textwidth]{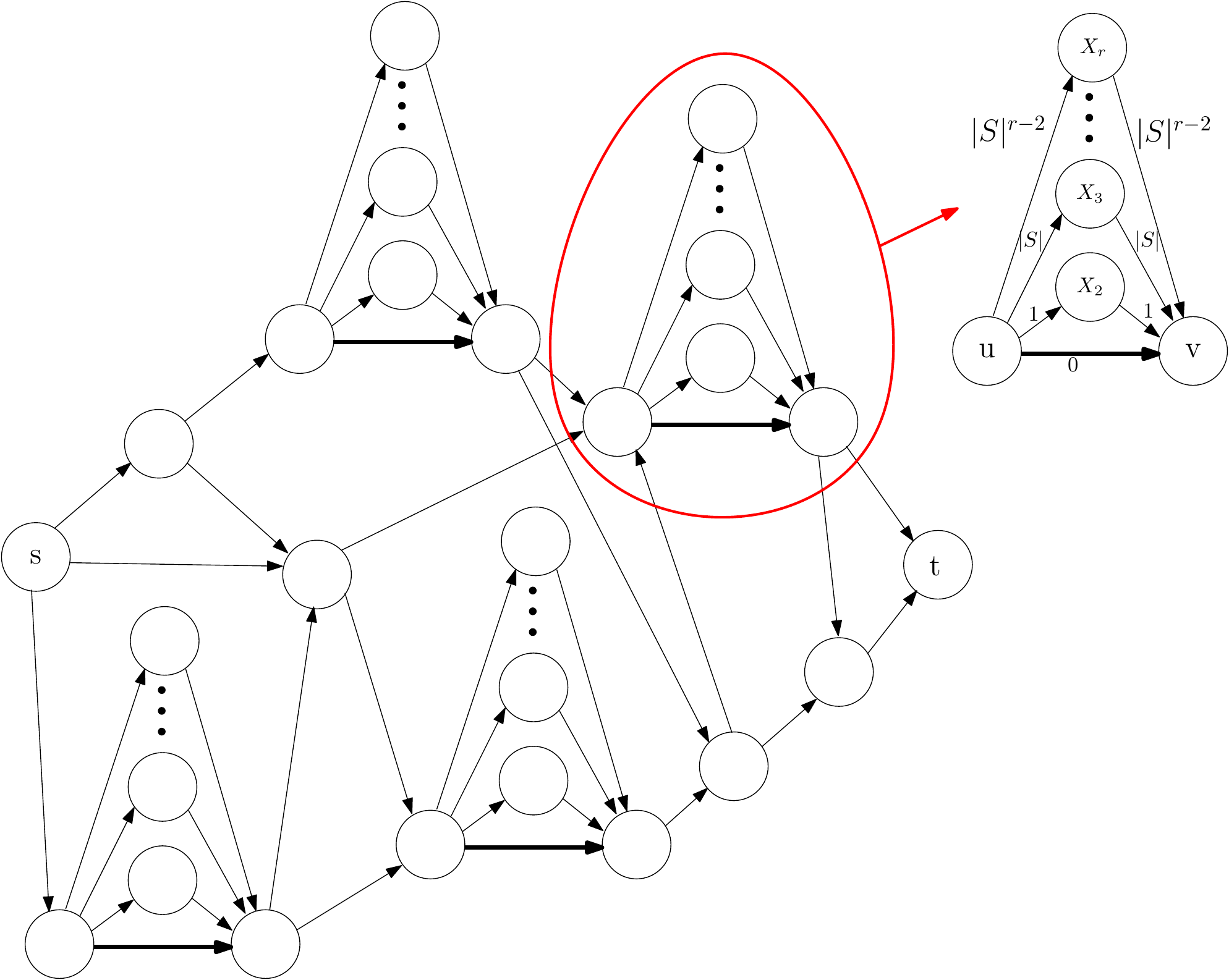}
    \caption{The construction of Proposition~\ref{prop:prop4}. The thick edges belong to $S$. In each gadget, all edge capacities are 1 and edge costs are as shown in the zoomed copy.}
    \label{fig:prop4}
\end{figure*}
\begin{itemize} 
    \item Graph $\tilde G$ is acyclic. For any node $u$ of $\tilde G$, $d(s,u)+d(u,t)=d(s,t)$. For any edge $(u,v)$ of $\tilde G$, $d(s,u)+w(u,v)+d(v,t)=d(s,t)$.
    \item A path from $s$ to $t$ in  $G$ is shortest if and only if it belongs to $\tilde G$. 
    \item Any path in $\tilde G$ is shortest between its end-nodes in $G$.
    \item If $v$ is reachable from $u$ in $\tilde G$, then all shortest paths from $u$ to $v$ in $G$ are contained in $\tilde G$.
\end{itemize}
Since we need only the subgraph of shortest paths for studying the shortest paths from $s$ to $t$, \emph{we assume $G=\tilde G$} in the follows, for the simplicity of notation.

First, assume that our goal is finding the maximal number of disjoint shortest paths. Let  us build flow network $N_1=(G,s,t,u_1)$ by assigning capacity $u_1(e)=1$ to each edge $e$ of $G$.
The following statement is shown in \cite[Section 4]{DBLP:conf/cscml/Dinitz21}.

\begin{proposition} 
   The maximal number of disjoint shortest paths is equal to the size of the maximal flow $f_{max}$ in $N_1$. The set of such paths can then be found by the flow decomposition of $f_{max}$.
\end{proposition}

Let us assume now that a \emph{set of sensitive edges $S \subseteq E$} is distinguished in $G$. Let us define flow network $N_2=(G,s,t,u_2)$ by assigning capacity $u_2(e)=1$ to each edge $e \in S$ and $u_2(e)=\infty$ to each other edge of $G$.

\begin{proposition}
   The maximal number of shortest paths disjoint at the edges in $S$ is equal to the size of the maximal flow $f_{max}$ in $N_2$. The set of such paths can then be found by the flow decomposition of $f_{max}$.
\end{proposition}

Recall that the case of node capacities can be easily reduced to that of edge capacities. Therefore, this and the following statements related to $S$ can be extended to the case of the \emph{set $S \subseteq V$ of sensitive nodes} in $G$.

Assume now that any edge of $S$ may be \emph{overloaded by at most two paths} going along it, and we look for the set of $r$ shortest paths minimizing the number of overloaded edges. Let us define flow network with edge-costs $N_3=(G,s,t,u_3,c_3)$ by taking $N_2$, assigning edge costs zero to all its edges, and for any edge $(u,v) \in S$, adding a new node $x_2$ and a pair of edges $(u,x_2)$ and $(x_2,v)$ of capacity 1 and of cost 1.

\begin{proposition}
   The set of $r$ shortest paths overloading any edge of $S$ by at most two paths going along it and minimizing the number of overloaded edges can be constructed by finding the min-cost flow $f_{mincost}$ of size $r$ in $N_3$ and applying to it the flow decomposition.
\end{proposition}

Now, assume that any edge of $S$ may be \emph{overloaded by any number of paths} going along it, and the objective is to minimize the maximum overload over all sensitive edges. That is, the loss of overloading even a single sensitive edge by $k$ paths is more than overloading all sensitive edges by $k-1$ paths, for any $k \ge 2$ (the prioritized loss). We look for the set of $r$ shortest paths minimizing the total loss of overloading the sensitive edges. 
Let us define flow network with edge-costs $N_4=(G,s,t,u_4,c_4)$ by taking $N_3$, and for any edge $(u,v) \in S$ and any $i: 3 \leq i \leq r$, adding a new node $x_i$ and a pair of edges $(u,x_i)$ and $(x_i,v)$ of capacity 1 and of cost $|S|^{i-2}$.
See Figure~\ref{fig:prop4} for illustration.

\begin{proposition}
\label{prop:prop4}
   The set of $r$ shortest paths minimizing the total prioritized loss of overloading sensitive edges can be constructed by finding the min-cost flow $f_{mincost}$ of size $r$ in $N_4$ and applying to it the flow decomposition.
\end{proposition}

    \remove{
\section{Concluding Remarks}
\label{s:renarks}
We have also investigated the $k$ near shortest trees problem not being aware of Sede{\~{n}}o{-}Noda and Gonz{\'{a}}lez{-}Mart{\'{i}}n \cite{DBLP:journals/eor/Sedeno-NodaG10}. We came with a more algorithmic solution that is less efficient, but may bring more understanding to the construction, we differ details to the full version.
}

\bibliographystyle{splncs04}
\bibliography{Reference}

\begin{thebibliography}{10}
\providecommand{\url}[1]{\texttt{#1}}
\providecommand{\urlprefix}{URL }
\providecommand{\doi}[1]{https://doi.org/#1}

\bibitem{DBLP:conf/csr/Akhmedov20}
Akhmedov, M.: Faster 2-disjoint-shortest-paths algorithm. In: Computer Science
  - Theory and Applications - 15th International Computer Science Symposium in
  Russia, {CSR} 2020, Yekaterinburg, Russia, June 29 - July 3, 2020,
  Proceedings. pp. 103--116 (2020)

\bibitem{DBLP:conf/icalp/BentertNRZ21}
Bentert, M., Nichterlein, A., Renken, M., Zschoche, P.: Using a geometric lens
  to find k disjoint shortest paths. In: 48th International Colloquium on
  Automata, Languages, and Programming, {ICALP} 2021, July 12-16, 2021,
  Glasgow, Scotland (Virtual Conference). pp. 26:1--26:14 (2021)

\bibitem{DBLP:conf/esa/Berczi017}
B{\'{e}}rczi, K., Kobayashi, Y.: The directed disjoint shortest paths problem.
  In: 25th Annual European Symposium on Algorithms, {ESA} 2017, September 4-6,
  2017, Vienna, Austria. pp. 13:1--13:13 (2017)

\bibitem{DBLP:conf/gis/ChondrogiannisB15}
Chondrogiannis, T., Bouros, P., Gamper, J., Leser, U.: Alternative routing:
  k-shortest paths with limited overlap. In: Proceedings of the 23rd
  {SIGSPATIAL} International Conference on Advances in Geographic Information
  Systems, Bellevue, WA, USA, November 3-6, 2015. pp. 68:1--68:4 (2015)

\bibitem{DBLP:conf/edbt/ChondrogiannisB17}
Chondrogiannis, T., Bouros, P., Gamper, J., Leser, U.: Exact and approximate
  algorithms for finding k-shortest paths with limited overlap. In: Proceedings
  of the 20th International Conference on Extending Database Technology, {EDBT}
  2017, Venice, Italy, March 21-24, 2017. pp. 414--425 (2017)

\bibitem{DBLP:conf/cocoon/DengGH18}
Deng, Y., Guo, L., Huang, P.: Exact algorithms for finding partial
  edge-disjoint paths. In: Computing and Combinatorics - 24th International
  Conference, {COCOON} 2018, Qing Dao, China, July 2-4, 2018, Proceedings. pp.
  14--25 (2018)

\bibitem{Dinic70}
Dinitz, Y.: Algorithm for solution of a problem of maximum flow in a network
  with power estimation. In: Soviet Math. Doklady. vol.~11, pp. 1277--1280.
  Springer (1970)

\bibitem{DBLP:conf/cscml/Dinitz21}
Dinitz, Y., Dolev, S., Kumar, M.: Polynomial time k-shortest multi-criteria
  prioritized and all-criteria-disjoint paths. In: Cyber Security Cryptography
  and Machine Learning - 5th International Symposium, {CSCML} 2021, Be'er
  Sheva, Israel, 2021, Proceedings. Lecture Notes in Computer Science, Springer
  (2021)

\bibitem{Eilam98}
Eilam{-}Tzoreff, T.: The disjoint shortest paths problem. Discret. Appl. Math.
  \textbf{85}(2),  113--138 (1998)

\bibitem{DBLP:journals/algorithmica/ErdmannL87}
Erdmann, M.A., Lozano{-}P{\'{e}}rez, T.: On multiple moving objects.
  Algorithmica  \textbf{2},  477--521 (1987)

\bibitem{DBLP:journals/corr/Jain21}
Gajjar, K., Jha, A.V., Kumar, M., Lahiri, A.: Reconfiguring shortest paths in
  graphs. CoRR  \textbf{abs/2112.07499} (2021),
  \url{https://arxiv.org/abs/2112.07499}

\bibitem{DBLP:conf/ijcai/GuoDLHSH18}
Guo, L., Deng, Y., Liao, K., He, Q., Sellis, T., Hu, Z.: A fast algorithm for
  optimally finding partially disjoint shortest paths. In: Proceedings of the
  Twenty-Seventh International Joint Conference on Artificial Intelligence,
  {IJCAI} 2018, July 13-19, 2018, Stockholm, Sweden. pp. 1456--1462 (2018)

\bibitem{DBLP:journals/algorithmica/HalperinLW00}
Halperin, D., Latombe, J., Wilson, R.H.: A general framework for assembly
  planning: The motion space approach. Algorithmica  \textbf{26}(3-4),
  577--601 (2000)

\bibitem{DBLP:journals/siamcomp/HopcroftW86}
Hopcroft, J.E., Wilfong, G.T.: Reducing multiple object motion planning to
  graph searching. {SIAM} J. Comput.  \textbf{15}(3),  768--785 (1986)

\bibitem{Jenn1997}
Jennings, J., Whelan, G., Evans, W.: Cooperative search and rescue with a team
  of mobile robots. In: 1997 8th International Conference on Advanced Robotics.
  Proceedings. ICAR'97. pp. 193--200 (1997)

\bibitem{DBLP:conf/soda/Lochet21}
Lochet, W.: A polynomial time algorithm for the \emph{k}-disjoint shortest
  paths problem. In: Marx, D. (ed.) Proceedings of the 2021 {ACM-SIAM}
  Symposium on Discrete Algorithms, {SODA} 2021, Virtual Conference, January 10
  - 13, 2021. pp. 169--178. {SIAM} (2021)

\bibitem{DBLP:conf/iros/MataricNS95}
Mataric, M.J., Nilsson, M., Simsarin, K.T.: Cooperative multi-robot
  box-pushing. In: Proceedings of {IEEE/RSJ} International Conference on
  Intelligent Robots and Systems, {IROS} 1995, August 5 - 9, 1995, Pittsburgh,
  PA, {USA}. pp. 556--561 (1995)

\bibitem{DBLP:conf/focs/PapadimitriouRST94}
Papadimitriou, C.H., Raghavan, P., Sudan, M., Tamaki, H.: Motion planning on a
  graph (extended abstract). In: 35th Annual Symposium on Foundations of
  Computer Science, Santa Fe, New Mexico, USA, 20-22 November 1994. pp.
  511--520. {IEEE} Computer Society (1994)

\bibitem{DBLP:journals/jct/RobertsonS95b}
Robertson, N., Seymour, P.D.: Graph minors .xiii. the disjoint paths problem.
  J. Comb. Theory, Ser. {B}  \textbf{63}(1),  65--110 (1995)

\bibitem{DBLP:conf/icra/RodriguezA10}
Rodr{\'{\i}}guez, S., Amato, N.M.: Behavior-based evacuation planning. In:
  {IEEE} International Conference on Robotics and Automation, {ICRA} 2010,
  Anchorage, Alaska, USA, 3-7 May 2010. pp. 350--355 (2010)

\bibitem{DBLP:conf/iros/RusDJ95}
Rus, D., Donald, B.R., Jennings, J.: Moving furniture with teams of autonomous
  robots. In: Proceedings of {IEEE/RSJ} International Conference on Intelligent
  Robots and Systems, {IROS} 1995, August 5 - 9, 1995, Pittsburgh, PA, {USA}.
  pp. 235--242 (1995)

\bibitem{DBLP:conf/icra/SmithEH08}
Smith, B.S., Egerstedt, M., Howard, A.M.: Automatic deployment and formation
  control of decentralized multi-agent networks. In: 2008 {IEEE} International
  Conference on Robotics and Automation, {ICRA} 2008, May 19-23, 2008,
  Pasadena, California, {USA}. pp. 134--139 (2008)

\bibitem{10.2307/2629312}
Yen, J.Y.: Finding the $k$ shortest loopless paths in a network. Management
  Science  \textbf{17}(11),  712--716 (1971)

\end{thebibliography}

\end{document}